\documentclass[
aps,%
12pt,%
final,%
notitlepage,%
oneside,%
onecolumn,%
nobibnotes,%
nofootinbib,%
superscriptaddress,%
noshowpacs,%
centertags]%
{revtex4}

\usepackage{graphicx}
\usepackage{dcolumn}
\usepackage{bm}
\usepackage{hyperref}

\usepackage{rotating}
\usepackage{array}
\usepackage{amssymb}
\usepackage{tikz}

\def\be{\begin{equation}}
\def\ee{\end{equation}}
\def\bea{\begin{eqnarray}}
\def\eea{\end{eqnarray}}

\def\ecal{\mbox{$\cal E\,$}}

\def\<{\langle}
\def\>{\rangle}

\begin{document}

\title{Self-similar analogues of Stark ladders: a path to fractal potentials}

\author{\firstname{E.}~\surname{Sadurn\'i}}

\email{sadurni@ifuap.buap.mx}
\affiliation{
Instituto de F\'isica, Benem\'erita Universidad Aut\'onoma de Puebla, Puebla, M\'exico
}%
\author{\firstname{S.}~\surname{Castillo}}
\affiliation{
Instituto de F\'isica, Benem\'erita Universidad Aut\'onoma de Puebla, Puebla, M\'exico
 }%


\begin{abstract}
We treat the eigenvalue problem posed by self-similar potentials, i.e. homogeneous functions under a particular affine transformation, by means of symmetry techniques. We find that the eigenfunctions of such problems are localized, even when the potential does not rise to infinity in every direction. It is shown that the logarithm of the energy displays levels contained in families that are analogous to Wannier-Stark ladders. The position of each ladder is proved to be determined by the specific details of the potential and not by its transformation properties. This is done by direct computation of matrix elements. The results are compared with numerical solutions of the Schr\"odinger equation. 
\end{abstract}

\maketitle

\section{Introduction}

There is an illustrious history behind the application of symmetry methods to solid state physics and associated crystalline structures. This field typically includes problems with exact and broken symmetries under lattice operations. Notorious examples among the latter entail the use of Hamiltonians which transform  in a very special way under a discrete group.

When Bloch oscillations were proposed \cite{bloch1928} several decades ago, interesting discussions were triggered by the possibility of experimental realizations, all of them connected to Wannier--Stark ladders in solids \cite{wannier1, wannier2, wannier3, zak1, zak2}. It was clear from the beginning that the transformation properties of the Hamiltonian for a particle in a crystal would lead directly to families of equispaced energy levels known as ladders. However, there seemed to be certain discord regarding the  position of each ladder, as well as the commensurability of levels for strong electric fields and band mixing. The issue was unequivocally related to the particular details of atomic potentials and lied beyond the transformation properties of the Hamiltonian under lattice operations. In this direction, it was a clever observation by Wannier \cite{zak3, wannier4} on the invariance of the {\it evolution operator\ }what led to undisputable results of time periodicity, with additional overall phases. A happy outcome of such discussions was the eventual observation of the physical effect \cite{mendez} and its artificial ramifications \cite{pertsch1999, wilkinson1996}. 

Old discussions and modern applications have left valuable mathematical tools: The kq-representation (or Zak transform), the Wannier functions and the Wannier transform. In this contribution we show that the techniques involved in the description of the aforementioned phenomena can be generalized to other families of discrete transformations, such as translations combined with scalings, rotations and shears: the affine group. We explore their consequences on energies and eigenfunctions using the properties of Hamiltonians with self-similar potentials. A notable example studied in the past is a fractal curve acting as a potential (the Mandelbrot curve), analyzed by Berry in \cite{berry}. 

We shall see in section \ref{sec1} that the spectrum is classified in families analogous to Stark ladders, but whose levels are exponential rather than linear functions of potential parameters. It shall be proved that wavefunctions describe bound states in all cases, with further relations to wavelets \cite{daubechies} of at most two parameters. In addition to these immediate observations, we exploit the properties of the Hamiltonian in order to compute its matrix elements in a suitable basis; with this result we identify the position of each (infinite) family of levels with the eigenvalue of a reduced matrix or Hamiltonian. In section \ref{sec2} we provide an exactly solvable example and in \ref{sec3}, a brief conclusion.

\section{The Schr\"odinger equation under affine transformations in one dimension \label{sec1}} 

Let us recall here that Stark ladders emerge through the translation properties of the Hamiltonian

\bea
H_{\scriptsize \mbox{Stark}} = \frac{p^2}{2m} + V_{\scriptsize \mbox{Periodic}}(x) + \ecal x.
\eea
Under a lattice translation $x \mapsto x + a$ the Hamiltonian transforms as $H_{\scriptsize \mbox{Stark}} \mapsto H_{\scriptsize \mbox{Stark}} + \ecal a$. The repetition of this transformation at the level of the stationary Schr\"odinger equation generates a spectrum given by

\bea
E_n = \varepsilon + n a \ecal , \qquad n \in \mathbb{Z}.
\eea
The origin of each ladder is determined by $\varepsilon$ and it depends crucially on the shape of the fundamental cell defined by $ V_{\scriptsize \mbox{Periodic}}(x)$. Fortunately, we can be sure that a gradual decrease of $\ecal$ should recover the usual band structure of energy levels, implying that $\varepsilon$ must be labeled by a band index as quantum number.

\begin{figure}[h!]
\begin{center}  \includegraphics[width=12cm]{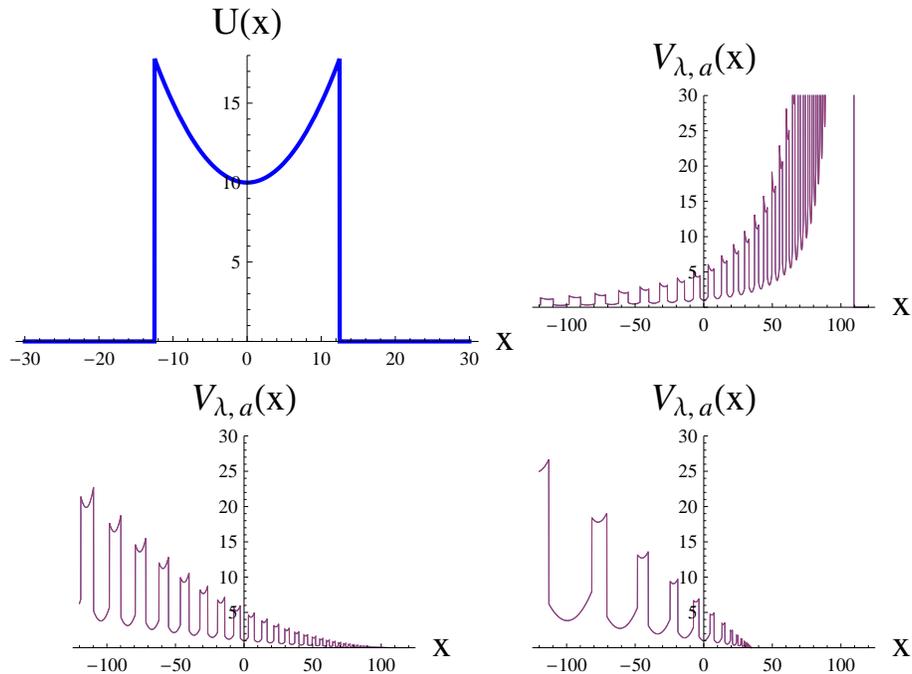} \end{center}
\caption{\label{fig:I.0} Left upper corner: Building block $U(x)$ with compact support, translated to the origin. The shape of $V_{\lambda, a}(x)$ resulting from $U$ as building block can be controlled by varying $\lambda$ and $a$, as shown in other panels.}
\end{figure}

Let us now consider a potential of the type 

\bea
H= \frac{p^2}{2m} + V_{\lambda} (x)
\label{scalablehamiltonian}
\eea
with the property
\bea
V_{\lambda} ( \lambda x) = \frac{V_{\lambda} (x)}{\lambda^2}, \qquad \mbox{for some} \quad \lambda.
\label{potential}
\eea
This type of restricted homogeneity ($\lambda$ is not arbitrary!) implies that under $x \mapsto \lambda x$, the Hamiltonian transforms as $H \mapsto H/\lambda^2$. Once again, the spectrum can be written  by successive transformations as

\bea
E_n = \varepsilon \lambda^{2n}, \qquad n \in \mathbb{Z}.
\eea
Evidently, the logarithm of the spectrum yields a Stark ladder

\bea
\log E_n = \log \varepsilon + 2n \log \lambda,
\eea
but the origin of each ladder --with the value $\log \varepsilon$-- is so far undetermined. Notable examples of $V_{\lambda}(x)$ include the Mandelbrot curve \cite{mandelbrot}, and the general form of such potentials can be given by expansions of the type

\bea
V_{\lambda} (x) = \sum_{n \in \mathbb{Z}} \lambda^{2n} U(\lambda^n x).
\label{series}
\eea
Mandelbrot's curve is obtained with the substitution $U(x) = \cos(\omega x)$ and the convergence of (\ref{series}) in this case is guaranteed  for certain values of $\lambda$ and $\omega$. There are other examples where the convergence of (\ref{series}) can be ensured for all values of $\lambda$ and almost all $x$; a function $U(x)$ with compact support away from the origin is one of them (see Fig. \ref{fig:I.0}).

The wavefunctions are now determined by the scaling properties of $H$. Performing the transformation once more at the level of the stationary Schr\"odinger equation

\bea
H \phi_n(x) = E_n \phi_n (x)
\eea
yields
\bea
\frac{1}{\lambda^2} H \phi_n(\lambda x) = E_n \phi_n( \lambda x)
\eea
or
\bea
H \left\{ \sqrt{\lambda} \phi_n(\lambda x) \right\} = E_{n+1} \left\{ \sqrt{\lambda} \phi_n(\lambda x) \right\}.
\eea
From here we see that if $\phi_n(x)$ is a square integrable function, so is

\bea
\phi_{n+1}(x) \equiv  \sqrt{\lambda} \phi_n(\lambda x).
\eea
Moreover, for some normalizable function $\phi(x,\varepsilon)$ corresponding to a ladder associated to $\varepsilon$, we can generate the complete family of functions with the formula

\bea
\phi_n(x) = \lambda^{n/2} \phi(\lambda^n x, \varepsilon).
\eea
A consequence of this formula is that the function $\phi(x,\varepsilon)$ can be identified as a mother wavelet of one parameter, since the orthonormality of the set $\left\{ \phi_n \right\}$ yields \footnote{We denote complex conjugation by a bar.}

\bea
\int_{-\infty}^{\infty} dx \bar \phi(\lambda^n x, \varepsilon) \phi(\lambda^m x, \varepsilon') = \lambda^{-n} \delta_{n,m} \delta_{\varepsilon, \varepsilon'}.
\eea
It should be noted that any method for the determination of $\varepsilon$ must be related to the shape of the building block, since the limits $\lambda \rightarrow 1$ or $\lambda \rightarrow 0$ recover gradually the features of $U(x)$ alone, but we postpone this discussion for the next section. Let us consider now the more general transformation

\bea
x_{n+1} = \lambda x_n + a
\eea
or 
\bea
x_n = \lambda^n x + \frac{1-\lambda^{n}}{1-\lambda} a, \qquad x_0 \equiv x.
\label{rule}
\eea
This affine transformation has the virtue of connecting the scale transformation with a lattice translation of period $a$ in the limit $\lambda \rightarrow 1$. A potential fulfilling partial homogeneity is now given by the expansion

\bea
V_{\lambda, a} (x) = \sum_{n \in \mathbb{Z}} \lambda^{2n} U(x_n)
\label{potentialV}
\eea
with $x_n$ given by (\ref{rule}). Properly stated, this means that

\bea
V_{\lambda, a} (\lambda x + a) &=& \sum_{n \in \mathbb{Z}} \lambda^{2n} U( x_{n+1})
= \sum_{n \in \mathbb{Z}} \lambda^{2(n-1)} U( x_{n})
= \frac{1}{\lambda^2} V_{\lambda, a} (x).
\eea
Since the Hamiltonian $H=p^2 / 2m + V_{\lambda, a}(x)$ has the same scaling properties as (\ref{scalablehamiltonian}), we can show again that the spectrum has the structure

\bea
E_n = \varepsilon(a) \lambda^{2n},
\eea
while the wavefunctions are now given by
\bea
\phi_n(x) =\lambda^{n/2} \phi(x_n, \varepsilon).
\label{trans}
\eea
Our duty now is to show that the value of $\varepsilon(a)$ comes indeed from the details of the building block $U(x)$.

\begin{figure}[h!]
\begin{center}  \includegraphics[width=12cm]{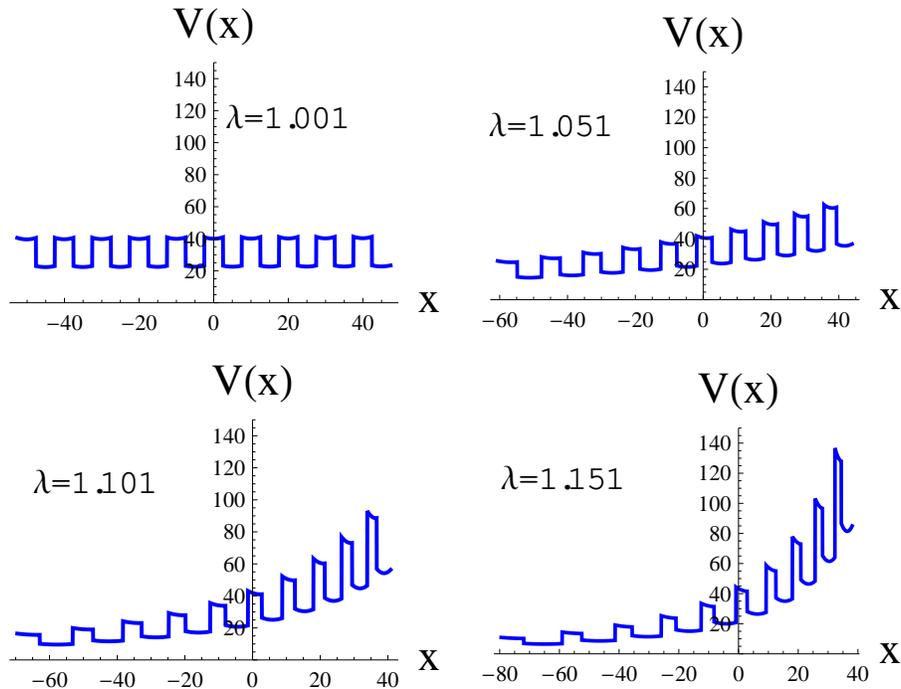} \end{center}
\caption{\label{fig:I.1.1} Potentials $V_{\lambda}(x)$ resulting from the building block $U(x)$ with compact support. The different values of $\lambda$ control the slope of $V$, as well as the distance between levels shown in figure \ref{fig:I.1}.}
\end{figure}

\begin{figure}[h!]
\begin{center}  \includegraphics[width=12cm]{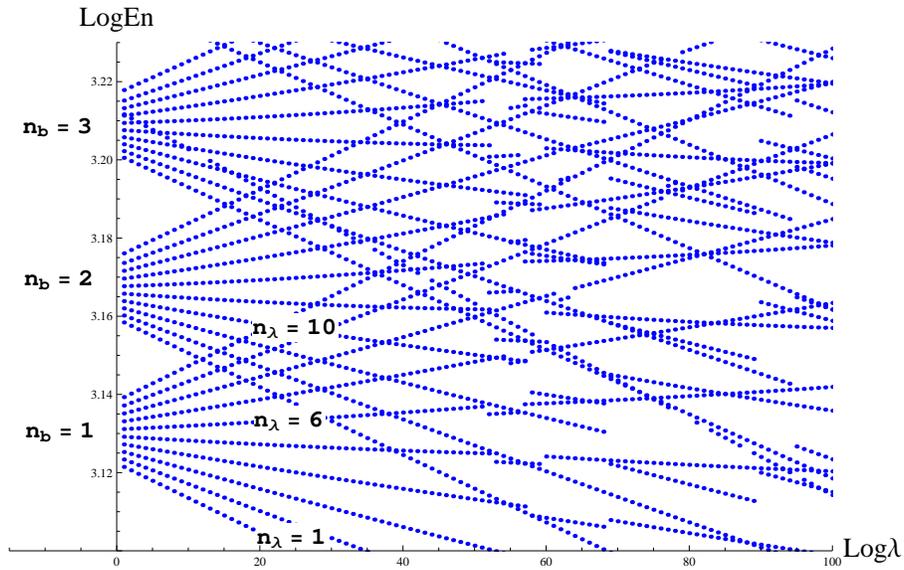} \end{center}
\caption{\label{fig:I.1} Energy levels of affine potential (\ref{potentialV}) in log-log scale. The similarity with a Stark ladder is notable; the position of each band is determined by $n_b$ in the limit $\lambda \rightarrow 1$, which produces a band structure corresponding to a lattice of period $a$. Within each ladder, the levels are denoted by $n_{\lambda}$.}
\end{figure}

\begin{figure}[h!]
\begin{center}  \includegraphics[width=12cm]{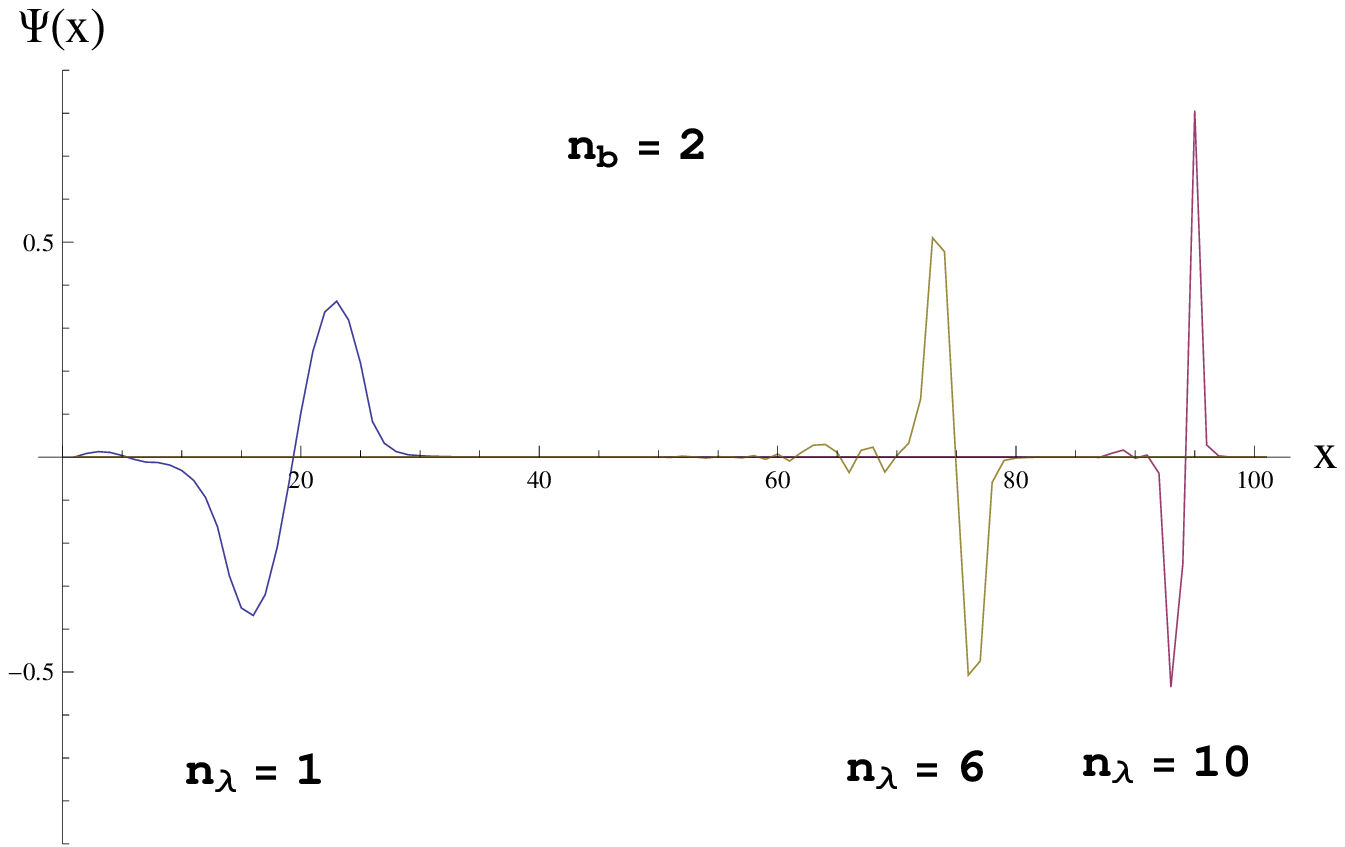} \end{center}
\caption{\label{fig:I.2} Eigenfunctions of a self-similar problem for levels $n_{\lambda} = 1,6,10$ in the same ladder $n_b=1$. The transformation properties between these functions are verified.}
\end{figure}

As an additional part of our study, we solve the eigenvalue problem posed by the potential $V_{\lambda}$ in Fig. \ref{fig:I.1.1} using other means. A numerical diagonalization of $H$ by the method of finite differences (with a very fine grain) reveals a linear trend of the energy levels in logarithmic scale. As it is evident, a ladder pattern emerges and the levels are grouped into well defined bands, except for finite size effects at the edges (see Fig. \ref{fig:I.1}). The corresponding wavefunctions are shown in Fig. \ref{fig:I.2}, where the shifts and rescalings relating the eigenstates of a ladder are displayed.

\section{Matrix elements and the determination of band indices}

\subsection{Finding the right basis}

We have seen that the transformation of $H$ leads naturally to the exponential part of the spectrum. The remainder, however, must be computed by a different method. Since there are many potentials with the same transformation rules under the discrete affine group, the position of each ladder must be determined by the specific shape of $U(x)$ in some fundamental cell. This suggests that a basis made of functions localized around the fundamental cell, should be capable of isolating the eigenvalue $\varepsilon(a)$ from the factor $\lambda^{2n}$.

First, we proceed to find such a set of localized waves inspired by periodic systems and Wannier functions. Then, by means of linear superpositions of our new basis, we shall construct the set of eigenfunctions of $H$.

\subsubsection{A scalable basis of localized functions}

The simplest way to find an orthonormal set of functions located predominantly around an interval is by finding the Wannier basis associated to a given periodic problem (for example, the one defined by the limit $\lambda \rightarrow 1$) whose period is precisely the length of the interval \footnote{Other options are less amenable. For instance, the periodic repetition of the set of functions for a particle in a box entails the use of fictitious Dirichlet conditions at the end of each cell, rendering the whole basis as incomplete.}. We denote such functions by

\bea
\chi_{n}^{k}(x) = \chi^{k}(x+ na) \equiv \< x  | k, n )
\label{x1}
\eea
(note the round bracket) where $a$ is the period of a cell, $k$ is the band index and $n$ is the site number. The transformation property (\ref{trans}) compels us to build a set of scalable functions $\left\{ \psi_{n}^{k} \right\}$ such that

\bea
\psi_{n}^{k}(x) = \lambda^{n/2} \psi^{k}(x_n), \qquad x_{n+1} = \lambda x_n + b,
\label{x2}
\eea
where we denote now by $b$ the translation parameter, as $a$ shall play a different role. 
To ensure completeness and orthonormality, we work now at the level of the inner product

\bea
\int_{-\infty}^{+\infty} \chi_{n}^{k} (x') \bar\chi_{n'}^{k'}(x') dx' = \delta_{kk'} \delta_{nn'},
\label{x3}
\eea
where the integral truly extends to $\mathbb{R}$, although $\chi_n$ may be localized around $x' = an$. Now we change variables according to

\bea
x = e^{x'} + \frac{b}{1-\lambda}, \qquad  x' = \log |x-\frac{b}{1-\lambda}|, \qquad \lambda \equiv e^a,
\label{x4}
\eea
and the inner product is transformed into

\bea
\int_{b/(1-\lambda)}^{+\infty} \frac{\chi^{k}\left[\log \left(\lambda^n|x-\frac{b}{1-\lambda}| \right) \right]  \bar \chi^{k'}\left[\log \left(\lambda^{n'}|x-\frac{b}{1-\lambda}| \right) \right]}{|x-\frac{b}{1-\lambda}|} dx = \delta_{kk'} \delta_{nn'}.
\label{x5}
\eea
Here the integration can be extended to $\mathbb{R}$ by noting the symmetry around $x=b/(1-\lambda)$, arriving thus at the following definition of orthogonal basis

\bea
\psi_n^{k} (x) \equiv \frac{\chi^k  \left[ \log \left( \lambda^n |x-\frac{b}{1-\lambda}| \right) \right]}{\sqrt{2|x-\frac{b}{1-\lambda}|}} = \< x| n, k \>,
\label{scalable}
\eea
which satisfies by construction the sought properties:

\bea
\int_{-\infty}^{\infty} dx \psi_{n}^{k} (x) \bar  \psi_{n'}^{k'} (x) dx = \delta_{kk'} \delta_{nn'},
\label{x6}
\eea
\bea
\psi_{n+1}^{k}(x) = \sqrt{\lambda} \psi_{n}^{k} (\lambda x + b).
\label{x7}
\eea
When $\lambda>1$, the function $\psi_n^k$ is predominantly localized at the interval $\lambda^n +b/(1-\lambda)<x<\lambda^{n+1} +b/(1-\lambda)$ if $\chi_n^k$ is localized around $ an<x'<an+a $. To this effect, it should be noted that the use of tempered functions for $\chi$ induces a similar temperance on $\psi$, i.e. if

\bea
\sqrt{|x|} \chi_{n}^{k}(x) \begin{array}{cc} \\ \longrightarrow \\  ^{\scriptsize |x| \to \infty} \end{array} 0, \qquad \forall n
\label{x8}
\eea 
then
\bea
\lim_{x \to \infty  \mbox{ \scriptsize or} \frac{ b}{1-\lambda} }  \sqrt{ \bigg| x-\frac{b}{1-\lambda} \bigg| \log \bigg| x - \frac{b}{1-\lambda} \bigg|} \quad \psi_{n}^{k}(x) =0, \qquad \forall n.
\label{x9}
\eea 
Moreover, if the Wannier function $\chi$ undergoes an attenuation stronger than exponential (e.g. gaussian envelopes) such as

\bea
e^{\alpha |x|} \chi_n^{k}(x) \begin{array}{cc} \\ \longrightarrow \\  ^{\scriptsize |x| \to \infty} \end{array} 0, \qquad \forall n, \quad \alpha>0,
\label{x10}
\eea
then we will have a strong localization of $\psi$ as well
\bea
\lim_{x \to \infty  \mbox{ \scriptsize or} \frac{ b}{1-\lambda} }   \bigg| x-\frac{b}{1-\lambda} \bigg|^{\alpha + 1/2}  \quad \psi_{n}^{k}(x) =0, \qquad \forall n,  \quad \alpha>0.
\label{x11}
\eea
\subsection{Matrix elements and eigenfunctions}

 The idea of defining a localized basis of scalable functions (\ref{scalable}) now comes into play: although they do not constitute an eigenbasis of $H$, our functions possess the property (\ref{trans}) and their localized nature ensures that the matrix elements of $H$ with different scale indices $n, n'$ gradually vanish as $n, n'$ become more distant. This is the cornerstone of band matrix approximations and one of the pillars of tight-binding models with successful applications to discrete systems \cite{sadurni2010, franco2013, atakishiyev1, atakishiyev2}. Let the eigenfunctions $\phi$ be given by the superposition 

\bea
\phi_n^k(x) \equiv \<x | n,k\>_E = \sum_{n',k'} U_{n,n'}^{k,k'} \psi_{n'}^{k'}(x)=\sum_{n',k'} U_{n,n'}^{k,k'} \<x|n',k'\>.
\label{x12}
\eea
It can be easily proved that (\ref{trans}) or its equivalent $\phi_{n+1}^{k}(x) = \sqrt{ \lambda} \phi_{n}^{k}(\lambda x + b)$ are fulfilled whenever (\ref{x7}) holds and the folllowing conditions on the expansion coefficients $U_{n,n'}^{k,k'}$ are met:

\bea
U_{n+1,n'}^{k,k'} = U_{n,n'-1}^{k,k'} \qquad \mbox{or} \qquad U_{n,n'}^{k,k'} = U_{n-n'}^{k,k'}.
\label{x13}
\eea
Meanwhile, orthonormality implies
\bea
\sum_{n',k'} U_{n-n'}^{k,k'} \bar U_{m-n'}^{q,k'} = \sum_{n',k'} \bar U_{n'-n}^{k',k} U_{n'-m}^{k',q} = \delta_{kq} \delta_{nm}.
\label{x14}
\eea
Since $H$ is assumed to be diagonal in $|n,k\>_E$, we also have

\bea
_E\< n,k| H | m,q \>_E = \delta_{nm} \delta_{kq} \varepsilon_k \lambda^{2n} = \sum_{n',m',k',q'} \bar U_{n-n'}^{k,k'} U_{m-m'}^{q,q'} \<n',k' |H|m',q' \>
\label{x15}
\eea
and we may invert this relation using (\ref{x14}), arriving to
\bea
\<n,k |H|m,q \> = \sum_{n',k'} U_{n'-n}^{k',k} \bar U_{n'-m}^{k',q} \varepsilon_{k'} \lambda^{2n'}. 
\label{x16}
\eea
This expression allows to show with little effort some useful properties involving the scale parameter:

\bea
\<n,k |H|m,q \> = \lambda^{2m} \<n-m,k |H|0,q \> =  \lambda^{2n} \<0,k |H|m-n,q \>
\label{x17}
\eea
\bea
\<n,k |H|m,q \> = \lambda^{n+m} \<\frac{n-m}{2},k |H|\frac{m-n}{2},q \> \equiv \lambda^{n+m} H_{n-m}^{k,q}.
\label{x18}
\eea
The $\lambda$ dependence in $H$ is now exhibited; the second factor in (\ref{x18}) does not change along the diagonals $n-m =s =$ constant, while the first factor increases as a power of $\lambda^2$. In matrix form, for fixed band indices $k,q$ we write

\bea
H^{k,q}=\left( \begin{array}{ccccccc}   &\ddots & \ddots   &  \ddots  &  & & \\ \cdots  &  \lambda^{-3} H_{1}^{k,q} &  \lambda ^{-2} H_0^{k,q}  & \lambda^{-1} H_{-1}^{k,q} & \cdots & & \\  & \cdots  & \lambda^{-1} H_{1}^{k,q} & H_0^{k,q}  &  \lambda H_{-1}^{k,q} & \cdots & \\  & & \cdots  & \lambda  H_{1}^{k,q} & \lambda ^2 H_0^{k,q}  &  \lambda ^3 H_{-1}^{k,q} & \cdots 
 \\  & &   & \ddots & \ddots  &  \ddots & 

  \end{array} \right).
\label{x19}
\eea
This means that it is possible to define a number operator $N$ with eigenvalues in $\mathbb{Z}$ and a shift operator $T$ such that \footnote{$T$ represents here an affine transformation and it could be expressed in terms of a translation and a squeezing operator, but we avoid this procedure for simplicity.}

\bea
N |n,k\> = n |n,k\>, \qquad T |n,k\> = |n+1,k\>, \qquad  T^{\dagger}|n,k\>= |n-1,k\>,
\eea
rendering 
\bea
H^{k,q} = \sum_{s= -\infty}^{\infty} H_{s}^{k,q} \left( \lambda^{N}  T^{s} \lambda^{N} \right) = \lambda^{N} \left( \sum_{s= -\infty}^{\infty} H_{s}^{k,q}  T^{s} \right) \lambda^{N}.
\label{x20}
\eea
This expansion corresponds indeed to a tight-binding Hamiltonian, where the index $s$ determines the range (or neighbor number) of the interaction and $H_s^{k,q}$ determines the intensity of the coupling (sometimes denoted as hopping parameter). The full Hamiltonian $H$ can be reconstructed as

\bea
H = \lambda^{N} H_0 \lambda^{N}, \qquad H_0 \equiv \sum_{s} T^s \left( \begin{array}{ccc}  H_{s}^{1,1} & H_{s}^{1,2} & \cdots \\ H_{s}^{2,1} & H_{s}^{2,2} & \cdots \\ 
\vdots & \vdots & \cdots
\end{array} \right)
\label{x21}
\eea
and the spectrum of the system can be obtained by a diagonalization of $H^{k,q}$ as in (\ref{x19}) (leading to eigenvalues which depend on $k,q$) followed by a final diagonalization in band indices $k,q$.

\subsection{Approximate form of the spectrum} In this language we can see clearly the main features of $E_{n,k}$ by applying a transformation $T$ to $H$. Using (\ref{x21}) and the commutation relations $\left[ T, N \right]= -T$, $\left[ T, H_0 \right]= 0$ one has

\bea
T H = \left( \lambda^{N-1} H_0 \lambda^{N-1} \right) T = \lambda^{-2} H T,
\label{x22}
\eea
and its successive application to $H |n,k \>_E = E_{n,k} |n,k\>_E $ leads to 
\bea
T^m H |n,k\>_E =  \lambda^{-2m} H T^{m}| n, k\>_E =  \lambda^{-2m} H | n + m, k\>_E    \nonumber \\ =  \lambda^{-2m} E_{n+m,k} | n + m, k\>_E =  E_{n,k} |n+m,k\>_E, 
\label{x23}
\eea
therefore $E_{n+m,k} = \lambda^{2m} E_{n,k}$, which is solved by $E_{n,k}= \lambda^{2n} \varepsilon_k$. Since the factor $\lambda^{2n}$ emerges independently of the explicit form of $H_0$, the only role played by $H_0$ is to fix $\varepsilon_k$. Disentangling the contribution of $H_0$ could be difficult in general situations for which $\left[ H_0, N \right] \neq 0$, implying that $H_0$ and $\lambda^{N}$ do not share eigenbases. However, $H_0$ is susceptible of nearest-neighbor approximations as we now discuss.

The overlaps which determine the range at which off-diagonal elements in (\ref{x19}) contribute are given by

\bea
\< s ,k |H| -s,q\> &=& \int dx \quad \bar \psi_{s}^{k} \left[ \frac{p^2}{2m} + \sum_{n \in \mathbb{Z}} \lambda^{2n} U(x_n) \right] \psi_{-s}^{q} \nonumber \\
&\approx& \int dx \quad \bar \psi_{s}^{k} \left[ \frac{p^2}{2m} + \lambda^{2s} U(x_s) + \lambda^{-2s} U(x_{-s}) \right] \psi_{-s}^{q} .
\eea
By virtue of (\ref{x9}), (\ref{x11}) and the discussion following (\ref{x7}), the further apart are $s$ and $-s$, the smaller is the overlap $\psi_s H \psi_{-s}$. The rate at which this overlap disappears obviously depends on the potential, which in turn is defined by the building block $U(x)$. For maximally localized functions, the limit case $H_{s}^{k,q} = \delta_{s0} \<0,k |H|0,q\>$ implies that $H_0$ is independent of $T$, i.e. $\left[ H_0 , N \right]=0$ and $H=  H_0 \lambda^{2N}$. Then the bases coincide $| n, k \>_E = | n, k \>$ and

\bea
 H |n, k \> = H_0 \lambda^{2N} |n,k\> = \lambda^{2n} H_0  |n,k\> = \lambda^{2n}\varepsilon_k  |n,k\>,
\eea
showing that $\varepsilon_k$ is directly the eigenvalue of $H_0$. Finally, this eigenvalue can be obtained by diagonalizing the simplified band-index matrix

\bea
\< 0,k |H| 0,q\> &=& \int dx \quad \bar \psi_{0}^{k} \left[ \frac{p^2}{2m} + U(x) \right] \psi_{0}^{q} .
\eea
Let us discuss another useful approximation to obtain the energy levels. We proceed by taking the periodic limit $\lambda \rightarrow 1$, which forces all levels in a ladder to coalesce and to form a continuous band. The center $\varepsilon_{k}$ of such a band determines the initial ($\lambda=1$) position of the eigenvalues. Then, the levels for arbitrary $\lambda$ can be obtained by appending $\lambda^{2n}$ to the formula. We may explain this in further detail as follows: by taking $\lambda \rightarrow 1$ in (\ref{rule}) with translation parameter $b$, we find that the matrix to be diagonalized is given by an overlap with the usual Wannier functions

\bea
\< n,k |H| m,q\> \rightarrow (n,k |H| m,q) &=& \int dx  \bar \chi_{n}^{k} \left[ \frac{p^2}{2m} + \sum_{n \in \mathbb{Z}} U(x+ n b) \right] \chi_{m}^{q} .
\eea
In this extreme situation, the operator $H$ has the same eigenvalues as $H_0$ and we know that such levels must be distributed in continuous bands -- each band is denoted by $k$. The operator $T$ becomes now a mere translation operator with continuous eigenvalue $e^{i \kappa}$, $\kappa \in \left[ -\pi,\pi \right]$. The general formula for the energies of a periodic system is always an expansion in periodic functions of $\kappa$:

\bea
\ecal_{\kappa,k} = \Delta_0^{k} + 2 \Delta_1^{k}  \cos \kappa + 2 \Delta_2^{k} \cos (2 \kappa) + 2 \Delta_3^{k} \cos (3 \kappa) + \cdots
\label{x24}
\eea
where each $\Delta_s^{k}$ is a Fourier coefficient. The center of each energy band then occurs at  a value of $\kappa$ for which $\ecal_{\kappa, k} = (\ecal_{\scriptsize\mbox{max}, k} - \ecal_{\scriptsize\mbox{min}, k})/2$. For $s_{\scriptsize \mbox{max}}=1$ (nearest neighbors) we have $\kappa = \pi/2$ and the center of the band is $\Delta_0^{k}$, leading to the following formula for the spectrum

\bea
E_{n,k} = \Delta_0^{k} \lambda^{2n}, \qquad \Delta_0^{k} = \frac{1}{2\pi} \int_{-\pi}^{\pi} d\kappa \quad \ecal_{\kappa, k}.
\eea
This expression proves to be very useful in understanding the numerical results shown in Fig. \ref{fig:I.1}. Each ladder opens symmetrically around a central energy, whose position is explained by the solutions of the periodic system. Moreover, the spectrum for $\lambda>1$ is {\it countable\ } and it has zero measure, leading to square integrable eigenfunctions. This explains the localization of waves depicted in Fig. \ref{fig:I.2}, despite the fact that the potential does not rise in the direction of the negative $x$ axis \footnote{This mechanism is related to the spectral measure of the system, but other examples of binding effects without increasing potentials have been extensively studied \cite{wignervonneumann, bittner}.}.

\section{An exactly solvable example \label{sec2}}

In previous sections we have shown that every scalable potential in one dimension leads invariably to a representation $H= \lambda^{N} H_0 \lambda^{N}$, $H_0$ depending on $U(x)$. We may propose a particular model for which $H_0$ allows exact calculations of energies and eigenfunctions.

To this end, let us assume a limited range for the couplings $H_s^{k,q}$. If the range is required to be bounded by a constant $\Delta^{k,q}$, i.e. $s_{\scriptsize \mbox{max}}\approx\Delta^{k,q}$, we may try a Bessel function of the first kind, whose asymptotic form \cite{watson} complies indeed with this requirement: 

\bea
H_s^{k,q} = \varepsilon \, i^s J_s(\Delta^{k,q}).
\eea
Here $\varepsilon$ is a constant with the dimensions of energy. For simplicity, let us take $\Delta^{k,q}$ as a diagonal matrix, so we can temporarily drop the band indices. The operator $H_0$ in (\ref{x20}) acquires now an appealing form

\bea
H_0 = \sum_{s=-\infty}^{\infty} \varepsilon i^s J_s(\Delta) T^s = \varepsilon \exp \left( \frac{\Delta}{2} (T + T^{\dagger}) \right)
\eea
where use has been made of the Jacobi--Anger expansion. The full Hamiltonian is now 

\bea
H=  \varepsilon \exp ( \log \lambda N )  \exp \left( \frac{\Delta}{2} (T + T^{\dagger}) \right) \exp ( \log \lambda N ).
\eea
Further simplification of this expression is attainable; our strategy is to cast the operators in a single exponential. Let us recall the following braiding identities:

\bea
\exp(X) \exp(Y) = \exp \left( e^{s} Y \right) \exp(X), \qquad \mbox{if}  \quad \left[X, Y \right]= s Y, \, s \in \mathbb{R}.
\label{y1}
\eea
In addition, if $\left[Y, Y^{\dagger} \right]= 0, \, X=X^{\dagger}$, then
\bea
\exp(X) \exp \left( Y - Y^{\dagger} \right) = \exp \left( e^s Y - e^{-s} Y^{\dagger} \right) \exp(X).
\label{y2}
\eea
Under these conditions, one has also
\bea
X + \lambda (Y + Y^{\dagger}) = \exp \left( \frac{\lambda}{s} (Y^{\dagger}-Y) \right) X \exp \left( \frac{\lambda}{s} (Y-Y^{\dagger}) \right)
\label{y3}
\eea
or taking the exponential in both sides
\bea
\exp \left[ X + \lambda (Y + Y^{\dagger} )\right] = \exp \left( \frac{\lambda}{s} (Y^{\dagger}-Y)\right) \exp(X)\exp \left( \frac{\lambda}{s} (Y-Y^{\dagger}) \right),
\label{y4}
\eea
and making use of (\ref{y2}), one can put the factors of $\exp(X)=\exp(X/2)\exp(X/2)$ to the left and the right to obtain
\bea
\exp \left[ X + \lambda (Y + Y^{\dagger} ) \right] = \exp(X/2) \exp \left[ \frac{2 \lambda \sinh(s/2)}{s} (Y + Y^{\dagger}) \right] \exp(X/2).
\label{y5}
\eea
The final step is to identify $X = s N$, $Y=T$, $s=2 \log \lambda$ and $\Delta = 4 \lambda \sinh( s/2 )/ s$, which yields

\bea
H = \varepsilon \exp \left\{ 2 \log \lambda \left[ N + \frac{\lambda \Delta}{2\lambda^2-2} \left( T + T^{\dagger} \right) \right] \right\}.
\label{y6}
\eea
This simple formula expresses the fact that a self-similar Hamiltonian can be written as {\it the exponential of a Wannier-Stark ladder,\ }where the eigenfunctions of $H$ are the same as those of the operator

\bea
H_{\scriptsize \mbox{Stark}} =  N + \frac{\lambda \Delta}{2\lambda^2-2} \left( T + T^{\dagger} \right),
\eea 
which turn out to be Bessel functions \cite{yellin, sadurni2013}. The energy levels are determined by $\varepsilon \lambda^{2n}$ and the eigenfunctions are combinations of Wannier functions of the form

\bea
| n, k \>_E = \sum_{q=1}^{\infty}\sum_{m = -\infty}^{\infty}  (-)^{n-m} J_{n-m} \left( \frac{\lambda \Delta^{k,q}}{2\lambda^2-2} \right) | m, q \>
\label{y7}
\eea
where we have restored the band index $k,q$ in $\Delta$. This closes our discussion.

\section{Conclusion and outlook \label{sec3}}

In this contribution we have shown that using an algebraic approach, one can treat self-similar problems with arbitrary building blocks as exponentials of Wannier--Stark ladders. The type of systems that we have treated could be considered as borderline cases of symmetry: their Hamiltonians are only {\it shape\ }invariant. Such operators are mapped to functions of themselves under the action of the affine group, rather than being Casimirs susceptible of Schur representation lemmas. Therefore, we rely now on non-diagonal representations of symmetries, i.e. the $T's$ must be regarded as shift operators; these shifts organize the spectrum in families dubbed ladders. We have further suggested a way to determine the position of each ladder based on the limit $\lambda \rightarrow 1$. Our discussions have led also to a particular Hamiltonian whose overlaps can be adjusted to yield an exactly solvable problem. 

Our line of reasoning allows us to speculate that general fractal forms whose similarity dimension coincides with the Hausdorff dimension, give rise to a linear dependence of the spectrum on such a parameter. This might help to furnish our solutions with a classification of otherwise irregularly interwoven energy levels in generic systems, i.e. a proper assignment of quantum numbers. In the near future, we plan to apply our results to the study of two-dimensional billiards with fractal boundaries.

\begin{acknowledgments}
We are grateful to CONACyT for financial support under project CB2012-180585. S.C. also wishes to thank CONACyT for {\it beca-cr\'edito\ }293535. 
\end{acknowledgments}

\section*{References}

\end{document}